\documentclass[10pt]{iopart}

\usepackage{iopams}

\begin{document}

\title[Finite temperature symmetry breaking BEC dark matter]{Bose-Einstein condensate dark matter phase transition from finite temperature symmetry breaking of Klein-Gordon fields}

\author{Abril Su\'arez and Tonatiuh Matos}

\address{Departamento de F\'isica, Centro de Investigaci\'on y de Estudios Avanzados del IPN,\\
                Av. IPN 2508, M\'exico D.F, M\'exico.}
\eads{\mailto{asuarez@fis.cinvestav.mx}, \mailto{tmatos@fis.cinvestav.mx}}

\begin{abstract}
In this paper the thermal evolution of scalar field dark matter particles at finite cosmological temperatures is studied. Starting with a real scalar field in a thermal bath and using the one loop quantum corrections potential, we rewrite Klein-Gordon's (KG) equation in its hydrodynamical representation and study the phase transition of this scalar field due to a $Z_2$ symmetry breaking of its potential. A very general version of a nonlinear Schr\"odinger equation is obtained. When introducing Madelung's representation, the continuity and momentum equations for a non-ideal SFDM fluid are formulated, and the cosmological scenario with the SFDM described in analogy to an imperfect fluid is then considered where dissipative contributions are obtained in a natural way. Additional terms appear compared to those obtained in the classical version commonly used to describe the $\Lambda$CDM model, i.e., the ideal fluid. The equations and parameters that characterize the physical properties of the system such as its energy, momentum and viscous flow are related to the temperature of the system, scale factor, Hubble's expansion parameter and the matter energy density. Finally, some details on  how galaxy halos and smaller structures might be able to form by condensation of this SF are given.
\end{abstract}

\pacs{98.80.-k, 95.53.+d, 67.85.Hj}
\vspace{2pc}
\submitto{\CQG}

\section{Introduction}
One of the main goals of cosmology is to explain the present state of the Universe on the basis of physical laws. Observations indicate that dark matter (DM) makes up an important fraction of the matter that conforms the Universe. The range of possible candidates have included from weak interacting massive particles to axions, just to mention some examples. The nature of DM appears as one of the most important questions of modern cosmology. Many models trying to answer this question involve scalar fields (SF). Scalar field dark matter (SFDM) models, in which dark matter is a bosonic particle with zero spin, have become a serious alternative to the standard cosmological model known as $\Lambda$CDM, \cite{b1}-\cite{b3}.

Up to now the theory of general relativity completely ignores quantum physical effects, but since the last century it has been studied how quantum effects can become relevant at high energies. In fact, they should dominate parts of the Universe where the density of matter can be large. In considering SFDM models, the evolution of the SF is usually treated at the classical level; however, quantum fluctuations may alter its potential. In the late Universe, $\hbar$ fluctuations may be harmless for most of the potentials, and it has been numerically shown how the energy density of the field is almost the same with or without quantum corrections, i.e., stable against one-loop corrections to the potential \cite{b30}-\cite{b32}.

Even though there exist regions for which quantum corrections of $\hbar$ can be dominant, we believe that, {\itshape a priori}, these corrections must be taken into account in any realistic model of
SFDM for the formation of structure, if this is to simulate a condensed system. This is due to the
fact of how long it takes the SF to leave the region where quantum corrections are important
(how long it takes for the breaking of symmetry to occur and possibly have a condensed
halo).

Even if it may not be clear from the potential how quantum effects can change the dynamics of the SF, it will be shown through the equations of motions (continuity and momentum equations, which will come from the KleinÐGordon (KG) equation) how important these quantum terms are appearing on the dissipative terms and the quantum potential in the momentum equations.

In this work, we consider one-loop contributions to the effective potential \cite{b4} ,and see how these affect the theory of SFDM through the dynamics of the hydrodynamical system of the KG equation, where in general it is known that the Klein-Gordon (KG) equation can be interpreted as the relativistic version of the Schr\"odinger equation describing the motion of a field, whose quanta are spinless particles.

The phenomenon of Bose-Einstein condensation (BEC) has now continued to be of great interest to theorists and experimentalists in different areas of physics. In recent years, analogies between different condensed matter systems and various branches of physics have been developed, see \cite{b5, b6}. In his work \cite{b7}, Ure\~na-Lopez pointed out how SF particles can form BECs at finite temperatures, work that had been carried out years before by Parker and Zhang \cite{b8, b9}. 

As an important conclusion, both pieces of work pointed out that a condensate can be considered relativistic when the temperature of the condensate is significantly larger than the mass of the boson, case that can be applied to this work. The cosmology of BECs has also been widely studied theoretically in \cite{b10} and in various numerical simulations in \cite{b11}.

The symmetry-breaking mechanism has been very useful for the study of phenomena associated  with BECs and phase transitions. The increase on the study of condensed matter systems through symmetry breaking has recently taken cosmology to point this way, and this is one of the motivations for this work.

In the cosmological context, symmetries can be gained when there are high temperature effects, as the temperature is lowered, spontaneous symmetry breaking (SSB) can give place to phase transitions both at local and global scales. In this scenario the evolution of the Universe will proceed in the same way as in the Hot Big Bang model, with the SF condensing out as temperature lowers.

In this work, one of the objectives is to study a cosmological system composed of a real SF as DM, this SFDM candidate can undergo a phase transition as the temperature of the thermal bath that it is in lowers. The SFDM particles decouple from the thermal bath while they are still relativistic, i.e., the decoupling temperature is much more higher than the mass of the SF particles. 

In analyzing such breaking of symmetry, a model with a $Z_2$ symmetry is taken, considering the case of the double-well potential with interaction and extended one-loop temperature corrections. As it will be confirmed throughout this work, the theory will be characterized by a dimensionless coupling constant $\lambda <<1$, such that the symmetries of the theory can be determined in a simple way only by the minimums of the scalar field potential $V(\Phi, T)$.

In section \ref{sec1}, the main basis of this work are given, the spontaneous symmetry breaking of the SF at finite temperature is discussed and the critical temperature $T_c$ of this breaking of symmetry through the scalar potential $V(\Phi,T)$ is obtained. Then, in section \ref{sec2}, a cosmological
version of a nonlinear Schr\"odinger equation at finite temperature analogous to the GrossÐ
Pitaevskii equation for BECs is derived.  

A continuity equation for the mass density and a momentum equation (analogous to Navier-Stokes's equation) from KG's equation are also obtained, with the objective of analyzing the explicit differences with the classic perspective. In this case, additional terms will appear for such equations, terms that turn to their common form in the case of classical limiting conditions. Possible interpretations of the results of the model are then argued in subsection \ref{sec3} and finally section \ref{seccon} is devoted to a small summary of the results of this work.

\section{Theoretical framework}\label{sec1}
\subsection{Hypothesis}\label{subsec1}

In this work, an analysis of the cosmological behavior of an SF as candidate for DM, which is made up of ultra-light bosons, $m\lesssim 1$eV, is made. These bosons will be described by a real SF, $\Phi$, with a $Z_2$ symmetry. The main idea is that at the beginning the Universe was in thermal equilibrium, and as the Universe expanded the SF cooled down with the rest of the particles.

The model then assumes that the interactions of the SF with ordinary matter can be completely neglected since the early Universe. From here that the field will be described by a quartic potential with thermal corrections, $T$, up to one-loop due to the auto-interactions of the SF particles themselves, which will be described by a self-interacting term $\lambda$. Then only the expansion of the Universe will keep cooling the SF. 

In this case, the possibility of using analogous systems to those in condensed matter physics (Bose-Einstein condensates) to analyze some problems in cosmology will be of great importance. The condition that will help us to have a relativistic condensate its such that its temperature should fulfill $T>>m$, condition which might be certainly fulfilled even today \cite{b7}.

Because it is expected that the Universe expands adiabatically, the temperature of the SFDM has to be always very similar to the temperature of the Universe itself. This assumption may or may not allow the SFDM particles to become non-relativistic at some point in the evolution of the Universe. This allows us to study a big range of masses for the SFDM particles, from $m=1$eV to $m=1\times 10^{-23}$ eV, including mass values that are of the order or well below the present temperature of the Universe $m\lesssim T\sim 10^{-4}$eV.

For simplicity and physical relevance, the theory will be characterized by a couple of parameters, the mass $m$ and a dimensionless coupling constant, that, as we will see later on will take very small values $\lambda<<1$. These are handled such that the symmetries of the theory will be determined in a simple way only through the minima of the SF potential $V(\Phi, T)$. Physical conceivable extensions to a non-linear Schr\"odinger equation in a Universe in expansion (generalization of Gross-Pitaevskii equation) will be studied, and the nonlinear terms of the theory, all of them depending on space and time, will also be considered.

\subsection{The one-loop potential}

To begin with, we consider a simple quantum field theory of a real SF whose Lagrangian is taken as
\begin{equation}
\mathcal{L}=\frac{1}{2}g^{\mu\nu}(\partial_{\mu}\Phi\partial_{\nu}\Phi)+P(\Phi),
\end{equation}
where $\Phi$ is a spin zero field and $P(\Phi)$ will be a quartic polynomial, so the solution describes a free field. Note that the Lagrangian will be invariant under the discrete symmetry transformation $\Phi\longrightarrow -\Phi$. We take the Friedman-Lemaitre-Robertson-Walker metric with sign convention, $a(\eta)(-1,1,1,1)$ with $a(\eta)$ the scale factor in conformal time, related to the cosmological time $t$ through $d/d\eta=a(d/dt)$.

From quantum field theory, it is known that the dynamics of an SF is governed by the KG equation. In this case, we add a first-order interaction potential $\phi$ (later on such a potential will
represent the gravitational potential, i.e., the trapping potential of our system)
$$\mathcal{L}_{int}=-\frac{m^2}{\hbar^2}\phi\Phi^2,$$
such that the KG equation in curved space-time ($\mathcal{L}+\mathcal{L}_{int}$) will be given by
\begin{equation}
\Box\Phi+V,_\Phi=2\frac{m^2}{\hbar^2}\phi\Phi
\end{equation}
where $\Box=-\frac{\partial^2}{\partial t^2}-3H\frac{\partial}{\partial t}+\frac{1}{a^2}\nabla^2.$

The auxiliary field $\phi$ is introduced to make reference to the gravitational potential that
appears in the perturbed KG equation which is used to describe structure formation through an
SF, so $\phi$ represents the gravitational potential in our system and will satisfy PoissonÕs equation \cite{b33}.

On the other hand, in order to make an analogy with the theory of condensed systems described by the GrossÐPitaevskii equation, an external trapping field $\phi$ is introduced so that it interacts with the SF (which only interacts gravitationally with the rest of the matter) at first order and which will confine our system (gravitational potential which is confined to the DM halo). As will be seen later on, for the classical theory of condensed systems in the momentum equation, a trapping potential,$V_{ext}$, is needed in order to confine the system. In this case, this will be given by the gravitational potential $\phi$.

A model having a $Z_2$ symmetry is taken, considering the easiest case of a double-well interacting potential given by,
\begin{equation}
  V_{\mathrm{c}}(\Phi)=\frac{\lambda}{4\hbar^2}\left(\Phi^2-\frac{m^2}{\lambda}\right)^2.
  \label{eq:V}
\end{equation}
To illustrate some of the features of finite temperature effects, we consider this potential extended to one-loop, so that the total effective potential with temperature corrections is,
\begin{equation}
 V_{\mathrm{T}}(\Phi, T)=V_{\mathrm{c}}(\Phi)+\frac{\lambda}{8\hbar^2}T^2\Phi^2 -T_0T^4
 \label{eq:VT}
\end{equation}
where $m$ is the mass parameter, $T$ is the temperature of the thermal bath, $T_0=\pi^2/90\hbar^2$ and $\lambda$ is the interaction parameter, this result includes both quantum and thermal contributions and are such that $c=k_B=1$ for convenience. This potential has a temperature-dependent term in addition to the zero temperature part in equation (\ref{eq:V}). The variation of potential (\ref{eq:VT}) with temperature determines the stage at which the symmetry breaking of the system appears. This temperature-dependent part receives a contribution from the SF itself, contribution that is taken into account by the second term on the right hand side (RHS) of  equation (\ref{eq:VT}) \cite{b4}. 

One of the main difficulties to build SF-based models is the choice of the potential. This work is focused on one-loop temperature contributions, because the choice of this potential opens the possibility for the SF to undergo a phase transition and this approach to symmetry breaking accounts for the effects of the ambient (thermal bath) in the corrections of the classical potential, equation (\ref{eq:V}). Of particular interest in the area of BoseÐEinstein condensation is the theoretical expectation that this phenomenon might occur by exactly undergoing a process of
symmetry breaking.

Laboratory BECs have been found to depend deeply on both the interacting parameter and
temperature of the trap (having a BEC at $T=0$ only in the ideal case), i.e., BECs are strong
interacting systems among their particles and surroundings; you cannot have a BEC without
the bosons interacting. Another cosmological motivation for the use of this potential is that
inside the SFDM model, the temperature corrections (excited, thermal states) have helped to
fit the rotation curves of LSB galaxies in an extraordinary way \cite{b34}.

This is not a unique extension, and higher corrections or other potentials can be taken into account, but for the purpose of this work, we stayed only with the lowest correction and the simplest possible interpretation. Higher corrections to the potential can be considered for future analysis, but such corrections have been shown to be even smaller and have less impact on the dynamics of the system.

At an early stage of the Universe, it is considered that this SF was in local thermodynamical equilibrium with its surroundings, this means that the temperature for the thermal bath was such that $m<<T_\rmi\sim 10^9$GeV, where $T_\rmi$ is the initial value assumed for the temperature at the beginning of the Hot Big Bang, \cite{b7}. When the temperature is high enough, one of the minima of the potential is $\Phi=0$. At this point, the SF density is equal to $\rho_\Phi\sim T_0T^4$, i.e, radiation dominates. The last term in (\ref{eq:VT}) will dominate only at very high $T$, so that as the SF cools down, $T^4$ will be very small and can be dropped out.

The critical temperature, $T_{\mathrm{c}}$, at which the symmetry breaking of the SF potential occurs is calculated. For that, the critical points of (\ref{eq:VT}) are obtained through the condition
\begin{equation}
\frac{\partial V_{\mathrm{T}}(\Phi,T)}{\partial\Phi} =\left(-\frac{m ^2}{\hbar^2}+\frac{\lambda}{\hbar^2}\Phi^2
 +\frac{\lambda}{4\hbar^2}T^2\right)\Phi=0.
\label{eq:cond}
\end{equation} 
In this case, it is important to note that the mass term has to have a negative sign in order that the phase transition of the scalar field can have a valid interpretation. In general the critical temperature is reached for values of $T$ for which the symmetries in the minima of $V(\Phi, T)$ are either gained or lost. From equation (\ref{eq:cond}), we can see that there are two critical points. The first can be found at $\Phi=0$. 

When the temperature is high enough, $T>>T_{\mathrm{c}}$, equation (\ref{eq:VT}) ensembles the behavior of a $\Phi^2$ potential, and this critical point is an absolute minimum, which corresponds to a 'false' vacuum. As the temperature lowers this minimum becomes a maximum and the point $\Phi=0$ will now turn to be an unstable maximum of the potential; the temperature at which these transitions occur is determined then by the critical temperature $T_{\mathrm{c}}$, which from equation (\ref{eq:cond}) is found to be
\begin{equation}
T^2_{\mathrm{c}}=\frac{4m^2}{\lambda}.
\label{eq:temp}
\end{equation}

So, above this critical temperature the potential contains a minimum like in a $\Phi^2$ regime and the SF keeps oscillating around this minimum, as the temperature lowers further than $T_{\mathrm{c}}$ this minimum becomes a maximum and the scalar field transits into the real new minimum of the potential, having two options to do so,
\begin{equation}
\Phi_{\mathrm{min}}=\pm\frac{1}{2}(T_{\mathrm{c}}^2-T^2)^{1/2}=\pm\alpha.
\label{eq:min}
\end{equation}
So the breaking of symmetry manifests itself once a minimum different from zero appears, in this case $\alpha$, which was obtained from equation (\ref{eq:cond}).

Since the theory must be constructed about a stable extremum of the potential, the ground state of the system will be around $+\alpha$ or $-\alpha$, and the $Z_2$ symmetry  will be broken once the ground state is chosen. If $\alpha_+$ positive is taken then the potential (\ref{eq:VT}) takes the form
\begin{equation}
V(\alpha_++\Phi)=\frac{1}{2}\mathfrak{m}^2\Phi^2+\left({\frac{\lambda}{2}}\right)^{1/2}\mathfrak{m}\Phi^3+\frac{\lambda}{4}\Phi^4,
\label{juan}
\end{equation}
where the positive mass $\mathfrak{m}$ in the minimum of the potential as $\mathfrak{m}^2=2m^2$ has been defined. After the SF passes through the breaking of symmetry, $T<<T_{\mathrm{c}}$, (and possibly a phase transition), the scalar potential is stabilized and the SF begins to oscillate around its minimum, at this stage the SF, $\Phi$, is supposed to take small values, so the terms that go as $\Phi^3$ and $\Phi^4$ in equation (\ref{juan}) can be neglected compare to those values for $\Phi^2$. 

Again we are left with a local $\Phi^2$-type potential. Once the SF choses its minima, we believe at this point that a Bose-Einstein condensate has been formed \cite{b33}.  At this point, it does not really matter which minima has the SF chosen, as the theory is equivalent. The same kind of studies for SFDM are being carried out with a U(1) symmetry.

From another point of view, from (\ref{eq:VT}) KG's equation, we have
\begin{equation}
\ddot\Phi+3H\dot\Phi-\left(\frac{\nabla^2}{a^2}-\frac{m^2}{\hbar^2}+\frac{\lambda}{4\hbar^2}T^2\right)\Phi+2\frac{m^2}{\hbar^2}\phi\Phi=0,
\label{eq:KG}
\end{equation}
which upon Fourier's transformation becomes
\begin{eqnarray}
\ddot{\Phi}_k+3H\dot{\Phi}_k+\left(\frac{k^2}{a^2}+\frac{m^2}{\hbar^2}-\frac{\lambda}{4\hbar^2}T^2+2\frac{m^2}{\hbar^2}\phi\right)\Phi_k=0
 \label{trans}
 \end{eqnarray}
Where $\Phi_k$ is the Fourier transform of the SF. From (\ref{trans}), the temperature can be parametrized with an index $n$ as $T^2=4n^2m^2/\lambda$. Equation (\ref{trans}) represents a harmonic oscillator with a damping term that goes as $3H$;  when this term results small there might be instability towards the growth of fluctuations. 

In this case the frequency of our system is given by $\omega^2=\frac{k^2}{a^2}+\frac{\lambda}{4\hbar^2}(T_{\mathrm{c}}^2-T^2)+2\frac{m^2}{\hbar^2}\phi_k$, which has to satisfy the condition $\omega^2<0$ in order for the SF fluctuations to grow. The thermal bath reaches the critical temperature $T_{\mathrm{c}}$ at $n=1$, see equation (\ref{eq:temp}); below this temperature, the fluctuations grow up till the temperature is very close to zero. From (\ref{trans}), parametrizing the KG equation with $n$, we have 
\begin{eqnarray}
\frac{k^2}{a^2}+\frac{m^2}{\hbar^2}-\frac{\lambda}{4\hbar^2}T^2+2\frac{m^2}{\hbar^2}\phi_k=\frac{k^2}{a^2}&+\frac{m^2}{\hbar^2}(1-n^2)+2\frac{m^2}{\hbar^2}\phi.
\end{eqnarray}

From here, it can be seen that when $n=1$ (we have reached the critical temperature and the SF has chosen a minima), the second term on the RHS of this equality vanishes and the fluctuations of the KG equation mimic the behavior of those in the standard model $\Lambda$CDM as seen in previous works by \cite{b14, b15}, satisfying the evolution equation
\begin{equation}
\ddot\Phi_k+3H\dot\Phi_k+\left(\frac{k^2}{a^2}+2\frac{m^2}{\hbar^2}\phi\right)\Phi=0,
\end{equation}
driven only by gravitational instability trough the gravitational potential $\phi$ and the size of the fluctuation determined through $k$. Bigger structures grow by hierarchy like in the CDM model, and grow exactly in the same way as it was shown in \cite{b33, b35}. 

This can be seen as an analogous way to the formation of BEC fluctuations, when the temperature of the universe is around $T_{\mathrm{c}}$ the SF breaks its symmetry and may condensate into BECs, forming SF 'drops' or in our case DM halos. In \cite{b11, b13}, the authors obtained that the the conditions for the collapse of the real scalar field into a massive object (oscillations) depended solely on the values for its mass $m\sim 1\times10^{-23}$eV and the interaction parameter $\lambda$ in accordance with what will be obtained later on throughout this work.

\section{Results}\label{sec2}

\subsection{Cosmological generalization of the non-linear Schr\"odinger equation and hydrodynamical representation}

In quantum field theory (and statistical mechanics), the quantity that changes through the phase transition and that characterizes  the difference between the two phases (before and after the breaking of symmetry) is known as the order parameter.

The order parameter of the free energy of a system that undergoes SSB during a phase transition has the following behavior: the minimum of the free energy that corresponds to a temperature $T>T_{\mathrm{c}}$ represents the unstable state; the phase transitions occurs at $T=T_{\mathrm{c}}$, once $T<T_{\mathrm{c}}$ one of the minima of the potential will represent the stable state of the system, which is just the behavior of the SF described in the last paragraphs. Then, in this case, $\Phi$ is the SF whose value will be analogous to the order parameter in the thermodynamical case. In the cosmological case, the finite temperature potential $V_{\mathrm{T}}$ mimics the free energy of the system associated with our SF, $\Phi$; for further reading see \cite{b17a}.

\subsubsection{The cosmological fluid}

To begin the analysis of the phase transition from a point of view analogous to that of the Ginzburg-Landau theory, we begin by writing the KG's equation (\ref{eq:KG}) in its hydrodynamical representation. In order to do so, the following ansatz is made:
\begin{equation}
\kappa\Phi=\frac{1}{2^{1/2}}[\Psi\mbox{exp}(-\rmi\mathfrak{m}t/\hbar)+\Psi^*\mbox{exp}(\rmi\mathfrak{m}t/\hbar)],
\label{eq:Phi}
\end{equation}
which involves the scale parameter $\kappa$ for further convenience in the dimensions, in this case for the dimensions of the SF, we have $[\Phi]=[M^{1/2}T^{-1}]$, referring $M$ and $T$ to dimensions of mass (energy) and time (length) respectively, so that the condition $\frac{1}{\kappa^2}|\Psi|^2=\frac{1}{\kappa^2}\Psi\Psi*=\hat\rho$, with $[\hat\rho]=[T^{-3}]$ can be fulfilled.

Then for KG's equation (\ref{eq:KG}), we have
\begin{equation}
 \rmi\hbar (\dot\Psi+\frac{3}{2}H\Psi)+\frac{\hbar^2}{2\,m}\Box\Psi+\frac{3\lambda}{2m\kappa^2}|\Psi|^2\Psi-m\phi\Psi\nonumber\frac{\lambda}{8m}T^2\Psi=0
\label{eq:Schrodinger}
\end{equation}
where we have used the notation $\dot{}=\partial/\partial t$. Equation (\ref{eq:Schrodinger}) is KG's equation in terms of the function $\Psi$ (order parameter) and temperature $T$ for an expanding Universe.
 
 In a non-expanding Universe with zero temperature contributions, $H=T=0$, and in the non-relativistic limit, equation (\ref{eq:Schrodinger}) becomes the Schr\"odinger equation with an external potential, in this case $\phi$. In this limit, this results in an analogous equation to that of Gross-Pitaevskii (being exactly the Gross-Pitaevskii equation for Bose-Einstein Condensates, when all the particles are in the ground state of the system, an approximate equation for the mean-field order parameter).
 
 The physical content of this generalized Gross-Pitaevskii like equation (\ref{eq:Schrodinger}) may be revealed by reformulating it as a pair of hydrodynamic equations, which are now derived. For this, $\Psi$ is expanded in its corresponding amplitude $\hat\rho$ and phase $S$ as
 \begin{equation}
  \Psi=\hat\rho^{1/2}\rme^{\rmi S},
 \label{eq:psi}
 \end{equation}
This is the so called Madelung representation and permits to understand the physical properties of Bose-Einstein condensation in a simpler way \cite{b16}-\cite{b18}. 

The Madelung representation is well known in the context of the ordinary linear Schr\"odinger equation, and it generalizes to the present cosmological situation without difficulty. Here, $\hat\rho=\rho/M_{\mathrm{T}}$ is interpreted as the number density of particles in the system which in this case are not yet in the condensed state (ground plus excited states) and which will be the rate between the mass density of particles $\rho=M_{\mathrm{T}}\hat\rho=M_{\mathrm{T}}\frac{1}{\kappa^2}|\Psi|^2=M_{\mathrm{T}}N/a^3$ and the total mass of particles in the system $M_{\mathrm{T}}$.  $N$ represents the total number of particles in the system and finally, $S=S(\vec{x},t)$, will be a function related to the velocity field as will be seen later on. 

Both $\hat\rho$ and $S$ are functions of time and position. To generalize our discussion, the following points are taken into account: (i) The velocity of the system depends upon the temperature of the particles (thermal velocity) and (ii) there are interactions between the particles that contribute to the potential energy of the system (characterized by $\lambda$).

In the classical approaches, Madelung's representation, equation (\ref{eq:psi}), is substituted into Gross-Pitaevskii's equation at zero temperature and the imaginary and real parts are worked with separately. Following the same path but this time at finite temperature, from equation (\ref{eq:Schrodinger}), we have, respectively,
 \begin{eqnarray}
\fl \dot{\hat{\rho}}+3H\hat{\rho}+\frac{\hbar}{m}(\hat{\rho}\Box S+\frac{1}{a^2}\nabla S\nabla\hat{\rho}-\dot{\hat{\rho}}\dot{S})=0\label{hidro1}\nonumber\\ 
\fl -\hbar\dot S-\frac{\hbar^2}{2ma^2}(\nabla S)^2+\frac{3\lambda}{2m}\hat\rho-m\phi+\frac{\lambda}{8m}T^2+\frac{\hbar^2}{2m}\left(\frac{\Box\hat{\rho}^{1/2}}{\hat{\rho}^{1/2}}\right)+\frac{\hbar^2}{2m}\dot S^2=0\label{hidro}
 \end{eqnarray}
In this case, we obtain the quantity 
\begin{equation}
U_{\mathrm{Q}}\equiv\frac{\hbar^2}{2m}\frac{\Box\hat\rho^{1/2}}{\hat\rho^{1/2}}
\end{equation}
which is a generalization (because of the curved space-time) of what is often called the 'quantum potential'. This term can work to prevent the BEC to collapse. 

When the number of SF particles in the system is not so high, the quantum potential can have a significative contribution to the phase transition, even though it is usually much smaller that the non-linear term, so it can be suppressed under certain conditions. Nevertheless, in this work, we keep such term throughout some parts of the analysis to obtain more general results. 
 
Taking the gradient of equation (\ref{hidro}), dividing by $a$ and using the following definition for the 'velocity field', 
 \begin{equation}
  \bi{v}\equiv\frac{\hbar}{ma}\nabla S
 \label{eq:vel}
 \end{equation}
we obtain
 \begin{eqnarray}
\fl\dot{\hat{\rho}}+\frac{1}{a}\nabla\cdot(\hat{\rho}\bi{v})+3H\hat{\rho}\left(1-\frac{\hbar}{m}\dot S\right)-\frac{\hbar}{m}(\hat{\rho}\dot{S}{\dot)}=0\label{eq:cont}\\
\fl\hat\rho\dot{\bi{v}}+\frac{\hat\rho}{a}(\bi{v}\cdot\nabla)\bi{v}=-\frac{\hat\rho}{a}\nabla\phi-\frac{\hat\rho}{a}\left[\nabla\left(-\frac{3\lambda}{2m^2}\hat\rho\right)\right]
-\frac{\hat\rho}{a}\nabla\left(-\frac{\hbar^2}{2m^2a^2}\frac{\nabla^2\sqrt{\hat\rho}}{\sqrt{\hat\rho}}\right)\nonumber\\
+\frac{\hbar}{m}\hat\rho\dot S(\dot{\bi{v}}+H\bi{v})
-\frac{\hbar^2}{4m^2a}\hat\rho\nabla\left(\frac{\ddot{\hat\rho}}{\hat\rho}\right)
+\frac{\hbar^2}{2m^2a}\hat\rho\nabla\left[\frac{1}{4}\left(\frac{\dot{\hat\rho}}{\hat\rho}\right)^2\right]
+\frac{\lambda}{4m^2a}\hat\rho T\nabla T\nonumber\\
-\frac{\hbar^2}{2m^2a}\hat\rho\nabla\left(\frac{3}{2}H\frac{\dot{\hat\rho}}{\hat\rho}\right)-H\hat\rho\bi{v}
 \label{eq:navier}
 \end{eqnarray}\label{eq:hydro}
which can also be written as
\begin{equation}
\dot{\bi{v}}+\frac{1}{a}(\bi{v}\cdot\nabla)\bi{v}=\bi{F}_{\phi}-\frac{1}{\hat\rho}\nabla p+\bi{F}_{\mathrm{Q}}+\frac{1}{\hat\rho}\nabla\sigma,
\label{vel2}
\end{equation}
where $\bi{F}_\phi=-\frac{1}{a}\nabla\phi$ is the force associated to the external potential $\phi$, and, as mentioned before, this potential is assumed to be the gravitational potential that satisfies the Poisson equation $\nabla^2\phi=4\pi G\rho$, with $\rho$ the energy density, $p=w\hat\rho^2$  is the pressure of the SF gas where $w=-3\lambda/2m^2a$ and $\bi{F}_{\mathrm{Q}}=-1/a\nabla U_{\mathrm{Q}}$ is the quantum force associated with the quantum potential. As we can see, the system can then be described in terms of a gas whose pressure and density are related by a bariotropic equation of state in the case of zero dissipative contributions \cite{b24,b28}.

With this, it is possible to define a local sound velocity in a formal way, varying such pressure with respect to the mass density of the 'fluid', $\rho=M_{\mathrm{T}}\hat\rho$, 
\begin{equation}
v_{\mathrm{s}}^2=\frac{\partial p}{\partial\rho}
\end{equation}
so a quantitative estimate for the magnitude of this velocity is given. As $p=\omega\hat\rho^2$, the velocity of sound will take the form
\begin{equation}
v_{\mathrm{s}}=\left(\frac{\omega\rho}{M_{\mathrm{T}}}\right)^{1/2}.
\end{equation}
Recall that in this case, $v_{\mathrm{s}}$ is not necessarily the physical sound velocity, but just a convenient parametrization.   

The function $\nabla\sigma$ is interpreted as an analogous to the flux viscosity expression that
appears in hydrodynamics; it contains not only terms which are gradients of the velocity but also of density and temperature, plus terms due to the expansion of the universe, i.e., it depends strongly on the dynamics of the system.

With this, $\nabla\sigma$ is defined as
\begin{eqnarray}
\fl\frac{1}{\hat\rho}\nabla\sigma= \frac{\hbar}{m}\dot S\dot{\bi{v}}+H\bi{v}\left(\frac{\hbar}{m} \dot S-1\right)+\frac{\lambda}{4m^2a}T\nabla T+\frac{1}{a}\zeta\nabla(\ln\hat\rho{\dot)}-\frac{\hbar^2}{4m^2a}\nabla\left( \frac{\ddot{\hat\rho}}{\hat\rho}\right)
 \label{eq:sigma}
 \end{eqnarray}
where the coefficient $\zeta$ is given by
\begin{equation}
 \zeta=-\frac{\hbar^2}{4m^2}\left[\frac{1}{\hat\rho a}\nabla\cdot(\hat{\rho}\bi{v})+6H-\frac{\hbar}{m}[3H\dot{S}+\frac{1}{\hat\rho}(\hat{\rho}\dot{S})\dot{}]\right]
 \label{eq:zeta}
 \end{equation}
and $\nabla(\ln\hat\rho{\dot)}$ as,
\begin{equation}
\fl\nabla(\ln\hat\rho{\dot)}=-\frac{1}{a}\nabla(\nabla\cdot\bi{v})-\frac{1}{a}\nabla[(\nabla\ln\hat{\rho})\cdot\bi{v}]+3H\frac{\hbar}{m}\nabla\dot S+\frac{\hbar}{m}\nabla\frac{1}{\hat{\rho}}(\hat{\rho}\dot{S}{\dot)}
 \label{eq:vrho1}
 \end{equation}
 
 As we can see from equation (\ref{eq:navier}), the total energy of the system will then have an additional  contribution that will come from the flux viscosity which apparently resists the expansion of the Universe. From equation (\ref{eq:sigma}), it is clear that the scale factor in the viscous term makes reference to the fact that the evolution of our system (DM halo) also resists the expansion of the Universe. 
 
To increase $\sigma$ means shortening $a$, so that at earlier times (smaller scale factor), the viscosity of the systems seems to be much larger; the viscosity has a bigger contribution to those SF haloes formed very early in time, so that, thanks to the contribution of the Hubble parameter in the friction of the system, it can have a better dissipation at higher temperatures (while the SF remains relativistic), possibly helping this way to have the right amount of collapsed halos, inhibiting the overpopulation of the substructure.
 
This model also incorporates a damping term that accounts for the finite temperature contribution in the system, which can always be obtained through the velocity distribution. Here, it should be mentioned that the temperature dissipation terms may result in interesting effects. In the typical case of a non-ideal fluid, the equation for the momentum commonly has a term of heat flux proportional to the temperature gradient  that goes as $\bi{F}=-K\nabla T$, where $K$ is related also to the temperature and is identified as a diffusion coefficient. From the RHS of equation (\ref{eq:sigma}), it can be seen that in this case there is also a term relating the gradient of the temperature to the temperature itself. The system (\ref{eq:cont}) and (\ref{eq:navier}) is completely equivalent to equation (\ref{eq:Schrodinger}). 

While taking into account only the first-order time derivatives (for simplicity of interpretation), the system (\ref{eq:cont}) and (\ref{eq:navier}) reads
 \begin{eqnarray}
  \dot{\hat{\rho}}&+&\frac{1}{a}\nabla\cdot(\hat{\rho}\bi{v})+3H\hat{\rho} =0\label{eq:contNR} \\ 
  \dot{\bi{v}}&+&\frac{1}{a}(\bi{v}\cdot\nabla)\bi{v}
  =\bi{F}_\phi-\frac{1}{\hat\rho}\nabla p+\bi{F}_{\mathrm{Q}}+\frac{1}{\hat\rho}\nabla\sigma\label{eq:navierNR}
    \end{eqnarray}\label{eq:hydroNR}
where $\nabla(\ln\hat\rho{\dot)}$ now reads
\begin{equation}
 \nabla(\ln\hat\rho{\dot)}=-\frac{1}{a}\nabla(\nabla\cdot\bi{v})-\frac{1}{a}\nabla[(\nabla\ln\hat{\rho})\cdot\bi{v}]
 \label{eq:vrho1NR}
 \end{equation}
 Thus, 
\begin{eqnarray}
\frac{1}{\hat\rho}\nabla\sigma=-H\bi{v}+\frac{\lambda}{4m^2a}T\nabla T
+\frac{1}{a}\zeta\nabla(\mbox{ln}\hat\rho)^{\dot{ }},
\label{eq:sigma2}
 \end{eqnarray}
where now we have,
\begin{equation}
 \zeta=-\frac{\hbar^2}{4m^2}\left[\frac{1}{a\hat\rho}\nabla\cdot(\hat{\rho}\bi{v})+6H\right]
 \label{eq:sig}
 \end{equation}

From equations (\ref{eq:cont}) and (\ref{eq:navier}), it is possible to notice that in the case of non expanding system ($H=0$)  and zero temperature conditions we obtain the already known classical hydrodynamical equations that allow us to describe Bose-Einstein condensates in the common way\\
-(1) Continuity equation:
\begin{equation}
\rho+\nabla\cdot(\rho\bi{v})=0
\end{equation}
-(2) Momentum equation:
\begin{equation}
\dot{\bi{v}}+(\bi{v}\cdot\nabla)\bi{v}=V_{\mathrm{ext}}-\frac{1}{\rho}\nabla p+\bi{F}_{\mathrm{Q}}+\frac{1}{\rho}\nabla\sigma,
\end{equation}
finding a consistent flux viscosity expression only depending on velocity gradients, proportional to $\nabla(\nabla\cdot\bi{v})$.

Clearly equations (\ref{eq:cont}) and (\ref{eq:navier}) differs from their counterparts in classical fluids mechanics, (1) and (2) given in the last paragraph, since now there are terms accounting for the expansion of the Universe and temperature contributions, even though these equations contain the classical equations of fluids in the classical limit and temperature equal to zero. 

If there is no space-time expansion, the viscosity flux for the BEC comes from gradients of temperature, velocity or/and the density as given by equations (\ref{eq:vrho1NR}) and (\ref{eq:sig}). This relation may in fact contain the whole information of the phase transition; then it will be interesting to follow the behavior of $T$ through the phase transition, classically reaching the condensed state once $T=0$. The main point here is that this phenomena is equivalent for a BEC on the earth and follows exactly the previous equations.

\subsubsection{Cosmological implications and phase transition}\label{sec3}

In \cite{b35}, it was previously found that when the condition $T<<T_{\mathrm{c}}$ was satisfied for the fluid, the SF density can be related to the density of the fluid by the relation $2\hat\rho=\rho_{\Phi_0}$, where $\rho_{\Phi_0}$ was identified as the density of the SF at the minimum. Taking this result, we can easily extend this study to account for temperature contributions; with the aid of equations (\ref{eq:min}) and (\ref{eq:Phi}) near the critical temperature of the breaking of symmetry $T_{\mathrm{c}}$, close to the minimum $\alpha$, we now find that the density of the SF fluid will take values around
\begin{equation} 
\hat\rho_{\mathrm{min}}=(T_{\mathrm{c}}^2-T^2)/4\kappa^2,
\label{eq:densT}
\end{equation}
so once $T<<T_{\mathrm{c}}$ the ground state will be even more occupied (note that in this case there still can be a quite numerous amount of particles in the excited states) and the breaking of symmetry and possibly the phase transition will occur once the conditions $T<T_{\mathrm{c}}$ or $\hat\rho>\hat\rho_{\mathrm{c}}$ are satisfied \cite{b19}.

In this case for the background velocity,  we have $\bi{v}_0=0$, so no gradients on velocity and temperature exist. In this limit from (\ref{eq:contNR}), we have
\begin{equation}
\frac{\partial\hat\rho_0}{\partial t}+3H\hat\rho_0=0,
\end{equation}
confirming that the density of the fluid will go as $\hat\rho_0= \hat\rho_0a^{-3}$ at late times, with $\hat\rho_0$ the density of the background. Thus from eq.(\ref{eq:densT}), the temperature of the fluid will decay as
  \begin{equation}
  T^2=T_{\mathrm{c}}^2-\frac{4\kappa^2\hat\rho_0}{a^3},
   \label{eq:solT0}
  \end{equation}
where $\hat\rho_0\sim 23\%\rho_{\mathrm{c}}$ and $\rho_{\mathrm{c}}\sim 4.19\times 10^{-11}$eV$^4$. If equation (\ref{eq:solT0}) is satisfied, it would then confirm, even in terms of the symmetry-breaking temperature, our common knowledge that satisfies the condition $T<<T_{\mathrm{c}}$ to have a phase transition and possibly a BEC.

There is yet another condition for a BEC to actually take place in the Universe. BEC is often claimed to be a phase transition without interaction. However, if all degrees of freedom are exactly free from any interaction, then there would be no transition from the difference states to the state of lowest energy. Therefore, there must be at least a small interaction which makes the transition possible.

In \cite{b20}, it was found that for interacting fields even with a small coupling constant, the mass of the halo can be large, with the mass of the structure given by $M\sim0.06\lambda^{1/2}\frac{m_{\mathrm{pl}}^3}{m^2}$. Using this result, if we want our SFDM to form structures as big as $10^{12}M_\odot$ with a scalar field mass around $m\lesssim 1$eV, an upper bound can be given to the coupling constant which has to have a value around
\begin{equation} 
\lambda^{1/2}\sim\frac{M}{0.06}\frac{m^2}{m_{\mathrm{pl}}^3}\lesssim 1.043\times 10^{-7},
\label{eq:acopl}
\end{equation}
where $m_{\mathrm{pl}}$ is Planck's mass. So indeed we have a small coupling parameter (and in general $\lambda$ will always have small values), confirming previous results \cite{b21,b26} .

Putting in numbers, from equations (\ref{eq:acopl}) and (\ref{eq:temp}), we have for the critical temperature of symmetry breaking,
\begin{equation}
T_{\mathrm{c}}\sim\frac{2m}{\lambda^{1/2}}\gtrsim 2\mbox{MeV},
\label{tval}
\end{equation}
so as we can see, the smaller the coupling parameter, the higher $T_{\mathrm{c}}$ will be \cite{b20, b22}.  So that if the possibility exists that the SF that forms the BEC lived in a thermal bath in the early Universe, from here we have that the symmetry breaking temperature corresponds to an epoch way before the radiation dominated era, but it turns out low enough to allow the SF to cool down with the rest of the matter. Following these results it might turn out possible to have well formed BEC massive halos since neutrino decoupling.

In a adiabatic expansion (cosmic evolution) in a Universe dominated by matter, the energy density evolves as $\rho\propto T^{3/2}$. As the temperature of matter is supposed to vary as $T_{\mathrm{m}}\propto a^2$, it follows that during the adiabatic evolution, the ratio between the temperature of the photons $T_{\gamma}$ and the temperature of matter varies as $T_{\gamma}/T_{\mathrm{m}}\propto a$. Therefore once the critical temperature of the boson (which will be given by (\ref{eq:solT0})), $T_{\mathrm{cr}}$, is below the critical temperature of the breaking of symmetry, equation (\ref{eq:temp}), BEC may occur at some stage during the cosmic evolution of the Universe.

In general, we also have the knowledge that BEC can be possible if the thermal de Broglie length exceeds the mean separation of particles of the bose gas, $\lambda_{\mathrm{dB}}\equiv[2\pi\hbar^2/(mkT)^{1/2}]>>r\equiv\hat\rho^{-1/3}$. In SFDM, we expect to have condensing temperates of the order of MeV or lower according to equation (\ref{tval}), taking into account that, in general, the mass of the SF will be small, then this will make the de Broglie length relatively large compared to the interparticle spacing of the SF particles, fulfilling then one of the conditions for BEC at cosmological temperatures.

Thus, in this paradigm, the DM halos can be born very early in time so that by recombination the halos are by then well formed. Then, if this paradigm is right we have to see big galaxies at high redshifts. If we expect that this SF forms a BEC, its temperature has to be of the order or below its critical temperature of symmetry breaking, i.e., once the temperature $T$ of the bosons goes below $T_{\mathrm{c}}$ of symmetry breaking, BEC may occur at some stage during the cosmological evolution of the Universe, with its dynamics governed by the equations obtained through this paper.

\subsection{Further comments}

\subsubsection{Quantum fluctuations and observational constraints.} As mentioned in the introduction, a bosonic particle that has been suggested to be the DM of the Universe is known as the axion. In general, if we initiate with metric perturbations, the initial fluid perturbations can be
separated into adiabatic and isocurvature components. Isocurvature perturbations are entropy
fluctuations.

The amplitude of these perturbations was tied to gravitational wave perturbations (tensor modes). While axions are a source of isocurvature perturbations and these play an important role in constraining axion models by their effects on the cosmic microwave background (CMB) \cite{b37}-\cite{b39} we did not consider them here because our SF is not an axion and we are only considering perturbations of the scalar type, so tensor perturbations will not have a considerable bearing on the formation of the condensed halo.

On the other hand, the SF that has been considered throughout this work is one of the kinds predicted by unification theories, string theories, etc, and is considered an independent quantum field from the inflaton, so that if isocurvature perturbations were present, these would be subdominant compared to adiabatic fluctuations. Such SF particles having masses down to $10^{-23}$eV. If these SF particles are not to overproduce isocurvature perturbations, the energy scale of inflation must be low and isocurvature contributions are not significant.

One can also think of this SF coupled to the inflaton before and during inflation, assumptions of this kind are not in the objectives of this paper. Throughout this paper, we suppose that SFDM only interacts gravitationally with the rest of the matter and that the thermal bath it is in comes only from the SFDM interactions itself.

If we were to suppose that this kind of DM is coupled to other matter since the beginning, their interactions would have to have an effect on the Big Bang nucleosynthesis (BBN), CMB distortions, probably affecting the different ratios of matter. The study of these perturbations and interactions in the formation of the halo was not in the aims of this paper. These kinds of couplings, while they maybe cosmologically important, are also beyond the scope of this work.

For the small SF masses that have been considered here, as can be seen from equation (\ref{tval}), the corresponding critical temperature of symmetry breaking is so high ($>>$MeV) that it is expected that the SF might be in its condensed state once equation (\ref{eq:solT0}) is fulfilled, i.e., the temperature of condensation is lower than the temperature of symmetry breaking, this occurring very early in time. Given this high critical temperature, we think that this field description can be used up to very early times.

Additionally, the effects for light SFDM particles with $m\lesssim 1$eV have been investigated in the acoustic peaks of the CMB. The results obtained in such a study also indicate that light SFDM particles might be present in a condensate state today. Then, it is concluded that this SF could mimic the effects of the standard $\Lambda$CDM model on the CMB spectrum \cite{b36,b21}.

While $\Lambda$CDM is non-relativistic at all times after decoupling, SFDM can be relativistic at early times, fulfilling the condition $T>>m$ (depending on the value of the mass of the SF that is being considered), before it becomes non-relativistic, behaving like $\Lambda$CDM dust at late times.

The transitions from going from one phase to the other have been shown to give important constraints on the SFDM model parameters, which are in accordance with the observation of the CMB and BBN \cite{b40}. This yields a mass constraint of $m>2.4\times 10^{-21}$eV and a constraint for the interacting parameter of the order of $\lambda<<1$ which are in accordance with the values we have taken throughout this paper.

As can be seen from equation (\ref{vel2}) in the case of zero dissipative contributions (if dissipative contributions are taken into account, the equation of state will also depend on $\sigma$ and a polytropic equation of state would no longer be satisfied), the polytropic equation of state goes as $p=\omega\rho^2$, where $\omega$ is of the order of $\omega\sim\frac{\lambda}{2m^2}$. In the case $\lambda<<m$, SFDM will behave like dust. In this case, it will evolve like $\lambda$CDM, with its background density going as $\hat\rho_0\propto a^{-3}$ as seen in \cite{b35,b40}.

In the case when the SFDM behaves like radiation, and $\lambda>>m$, in this limit, the equation of state will keep the form $p=\frac{3\lambda}{2m^2}\hat\rho^2$ (equation (\ref{vel2})). These cases confirm once again that in the case $\lambda=0$, SFDM will just behave like the standard $\Lambda$CDM. As can be seen from \cite{b35,b40}, these two cases have important results for this model.

From another point of view in \cite{b40}, it was shown that the transition of the behavior of SFDM from relativistic to non-relativistic $\Lambda$CDM has to occur early enough to be in agreement with the redshift of the radiation-matter equality ($z_{eq}$) epoch found in the CMB. They found that to keep this equality untouched, SFDM should be into the dust phase before or at least at $z_{eq}$.

Another result found by Bohua et al and independently by Briscese was that if SFDM were relativistic, it could contribute to $N_{eff}$ as an extra relativistic ingredient, putting constraints on the SFDM particles once again. Then they noted that SFDM could provide an explanation about the discrepancies between the $N_{eff}$ measured from the CMB and BBN \cite{b26,b40}.

With respect to the formation of the cosmic structure, if SFDM were very light, $m<1$eV, coherent oscillations of the field could lead to the suppression of clustering at certain scales, making it distinguishable from the $\Lambda$CDM \cite{b41}. Using the hydrodynamical representation (through equations (\ref{eq:cont})) and (\ref{eq:navier})) but without temperature contributions, it was found in \cite{b35} that if SFDM is treated as imperfect fluid with quantum potential contributions, an effective speed (analogous to the speed of sound) can be found, $v_q^2=\frac{\hbar^2k^2}{4a^2m}$, so that if we want no suppression in order for the perturbations to grow, then the condition $v_q<<1$ should be satisfied, defining scales such that the condition $k<<am$ should be accomplished.

\section{Conclusions}\label{seccon}
 
One of the objectives of this work was to obtain an extension to the possibility that scalar fields could contribute to the dark matter component of the Universe. In the search of relating this cosmological model composed of a real SF as DM with a system formed through BECs, a great number of characteristic features were found.

In particular, a couple of generalizations to a non-linear Schr\"odinger equation (\ref{eq:Schrodinger}) were obtained which in principle can be allowed on systems undergoing Bose-Einstein condensation at finite temperature. This relativistic non-linear Schr\"odinger equation (\ref{eq:Schrodinger}) is still a proposal, and the possibility of other forms cannot be excluded, depending on the form of the potential.

One of the biggest challenges that modern cosmology has is that of structure formation. To try and give a possible explanation, we gave the physical characteristics describing our system. For this, Madelung's representation was used (\ref{eq:psi}) and the temperature- and scale- factor-dependent hydrodynamical equations for the density and momentum were obtained, describing a gas whose density and pressure are related by a barotropic equation of state in the case of zero dissipative contributions, equations (\ref{eq:cont}) and (\ref{eq:navier}), and which could account in the dynamics of formation for DM halos in accordance with previous results obtained in \cite{b24, b25}. 

We conclude in a natural way, starting with Klein-Gordon's equation in a curved space-time (FLRW metric), that DM cannot be considered as an ideal gas. Instead, we treated it as a non-ideal fluid, i.e.,  with matter strongly correlated conductivity effects at finite temperature (possibly related with $\lambda$) and a dissipative effects through $\sigma$, equation (\ref{eq:sigma}). Under these approximations, the effects of the cosmological expansion are taken into account in the equations of motion for the fluid, making it able to follow the phase transition through the behavior of $T$. These generalizations may not be easy to study at all, but we where interested in searching how far we could go on analyzing such systems.

Making the classical approximations worked perfectly in accordance with the theory of non-viscous ideal fluids. The simplicity of our approaches does not aim to claim any concluding results whatsoever, but certainly may help to interpret the influence of the cosmological viscous terms in the evolution of structure formation. We mentioned some of the observational implications and constraints that can be obtained from the SFDM \cite{b40} in comparison with this analysis. As it is an important theme, we wrote a small subsection to summarize its implications.             

We expect that during the phase transition of condensation, when the thermal temperature of the bosons becomes of the order or lower to the critical temperature of symmetry breaking, $T_{\mathrm{c}}$, the gas condensates forming galaxy haloes, (\ref{eq:solT0}), see also \cite{b27}. For this we have given a lower limit for the critical temperature of  the breaking of symmetry, (\ref{tval}), below which Bose-Einstein condensation of our SF might occur. This implies that if the scalar field dark matter model is right and that condensation is actually a fact we have to see well formed galaxies haloes at big redshifts and all galaxies must have very similar features, because all of them form in a very short period of time under the same conditions.

\ack{This work was partially supported by CONACyT M\'exico under grants CB-2009-01, 132400, CB-2011, 166212 and I0101/131/07 C-234/07 of the Instituto Avanzado de Cosmologia (IAC) collaboration (
www.iac.edu.mx/).}

\end{document}